\newcommand{\be}{\begin{equation}}\newcommand{\ee}{\end{equation}}
\newcommand{\bea}{\begin{eqnarray}}\newcommand{\eea}{\end{eqnarray}}
\newcommand{\nn}{\nonumber}\newcommand{\p}[1]{(\ref{#1})}
\newcommand{\lb}[1]{\label{#1}}

\def\sfrac#1#2{{\textstyle\frac{#1}{#2}}}

\documentclass[12pt]{article}
\usepackage{amsmath,amssymb}

\begin{document}

\begin{center}
{\Large\bf
Supersymmetry at BLTP: how it started \\
\vspace{0.3cm}

and where we are}
\vspace{1.5cm}

{\large\bf
E.A. Ivanov} \\
\vspace{1cm}

 {\it Bogoliubov  Laboratory of Theoretical Physics, JINR,
141980 Dubna, Russia} \\
{\tt eivanov@theor.jinr.ru}

\end{center}
\vspace{1cm}

\begin{abstract}
\noindent This is a brief review of the BLTP activity in supersymmetry
initiated by V.I. Ogievetsky (1928-1996) and lasting for more than 30 years. The
main emphasis is made on the superspace geometric approaches and unconstrained
superfield formulations. Alongside such milestones as the geometric formulation
of $N=1$ supergravity and the harmonic superspace approach to extended
supersymmetry, I sketch some other developments largely contributed by the
Dubna group.
\end{abstract}

\setcounter{page}{1}
\section{Introduction}
Supersymmetry (SUSY) is a remarkable new type of relativistic symmetry which
combines into irreducible multiplets the particles with different spin and
statistics: bosons (integer spins, Bose-Einstein statistics) and fermions
(half-integer spins, Fermi-Dirac statistics). Since it transforms bosons into
fermions and vice-versa, the corresponding (super)algebras and (super)groups
involve both bosonic and fermionic generators. To avoid a contradiction with
the fundamental spin-statistics theorem, the fermionic generators should obey
the anticommutation relations in contrast to the bosonic ones which still
satisfy the commutation relations. Correspondingly, the group parameters
associated with the fermionic generators should be anticommuting (Grassmann)
numbers.

The actual interest in supersymmetries arose after the appearance of the papers
\cite{susy}-\cite{susy2} where self-consistent fermionic extensions of the
Poincar\'e algebra were discovered and their field-theoretic realizations were
found. The simplest ($N=1$) Poincar\'e supersymmetry, besides the standard
Poincar\'e group generators $P_m, L_{[m,n]}$ ($m, n = 0,1,2,3$, $P_m$ being the
4-translation generators and $L_{[m,n]}$ Lorentz group ones), involves the
fermionic Weyl generators $Q_\alpha, \bar Q_{\dot\alpha}\; (\alpha, \dot\alpha
= 1,2)$ which transform as $(1/2,0)$ and $(0,1/2)$ of the Lorentz group and
satisfy the following anticommutation relations: \be \{Q_\alpha,
\bar Q_{\dot\alpha}\} = 2(\sigma^m)_{\alpha\dot\alpha} P_m\,, \quad \{Q_\alpha,
Q_{\beta}\} = \{\bar Q_{\dot\alpha}, \bar Q_{\dot\beta}\} =0\,, \quad
(\sigma^m)_{\alpha\dot\alpha} = (1, {\vec{\sigma}})_{\alpha\dot\alpha}\,.\lb{1}
\ee $N>1$ extended supersymmetry involves $N$ copies of the fermionic
generators, each satisfying relations \p{1} \be \{Q_{\alpha}^{i}, \bar
Q_{\dot\alpha\,k}\} = 2 \delta^i_k (\sigma^m)_{\alpha\dot\alpha} P_m\,, \quad
\{Q_{\alpha}^{i}, Q_{\beta}^{k}\} = \{\bar Q_{\dot\alpha\,i}, \bar
Q_{\dot\beta\,k}\} =0\,.\lb{2} \ee
Here $i = 1, \ldots N$ is the index of a
fundamental representation of the internal automorphism symmetry (or
R-symmetry) group $U(N)\,$.

The possibility to achieve a nontrivial junction of internal symmetry with the
Poincar\'e symmetry by placing the fermionic generators into nontrivial
representations of the internal symmetry and thus to evade the Coleman-Mandula
theorem \cite{CM} is one of the remarkable new opportunities suggested by
supersymmetry. Nowadays it has a lot of theoretical manifestations and
applications, in particular, in String Theory. Another nice new feature follows
directly from relations \p{1}and \p{2}. Since the anticommutator of global
supersymmetry transformations produces a shift of $x^m$ ($P_m =
-i\frac{\partial}{\partial x^m}$), it is clear that the anticommutator of two
{\it local} supersymmetry transformations inevitably produces a {\it local}
shift of $x^m$. The gauge theory of local $x^m$ translations (or
R$^4$-diffeomorphisms) is the Einstein gravity. Hence, any theory invariant
under local supersymmetry transformations should include gravity. Since the
generators $Q_\alpha, \bar Q_{\dot\alpha}$ carry the spinor index of Lorentz
group, the associated gauge fields should be, first, fermions, and, second,
carry an extra vector index $m$, i.e., be represented by the Rarita-Schwinger
field $\psi^\alpha_m, \bar\psi^{\dot\alpha}_m\,$. So these massless gauge
fields should carry spin 3/2 (or helicity $\pm 3/2$ on the mass shell) and
form, together with the graviton $h_{mn}$, an irreducible supermultiplet (in
the general case of local $N$ extended supersymmetry this supermultiplet
contains more fields, with a nontrivial assignment with respect to the
R-symmetry group). Such an extension of gravity is the supergravity theory. By
definition, it is the gauge theory of linearly realized local supersymmetry and
as such it was discovered in \cite{SG} \footnote{First gauge theory of $N=1$
Poincar\'e supersymmetry, with the latter being nonlinearly realized as
spontaneously broken symmetry, was constructed in \cite{VSa}.}. Supergravity
theories are the only possible self-consistent field theories of an interacting
spin 3/2 field (with a finite number of gauge fields).

The discovery of supersymmetry at the beginning of the seventies was, to some
extent, an expected event for Victor Isaakovich Ogievetsky. This was one of the basic
reasons why the pioneering papers \cite{susy}-\cite{susy2} received a quick
enthusiastic respond in the group of theorists at LTP concentrating around
him (later on, Sector ``Supersymmetry'' headed by V.I. Ogievetsky for a long time).

In the sixties, V.I. Ogievetsky and I.V. Polubarinov put forward a new viewpoint on the
gauge fields (which on their own were a rather exotic concept at that time)
based on the so-called ``spin principle'' \cite{OP}-\cite{Lec}. They introduced
an important notion of the spin of an interacting field and argued that the
gauge invariance was just the device to ensure some massless interacting fields
to have a definite spin. They showed that requiring a massless vector field to
have spin 1 uniquely leads to Yang-Mills theory, while requiring a massless
tensor field $h_{mn}$ to possess spin 2 in interaction (actually with an
admixture of spin 0) yields Einstein theory.

In lectures \cite{Lec} Ogievetsky and Polubarinov posed a question about the
existence of the theory of interacting massless spin-vector field, such that
the latter carried the definite spin 3/2 in interaction. In other words, they proposed to
search for a theory in which the Rarita-Schwinger field played a role of a
gauge field, with the corresponding gauge invariance being intended to
eliminate a superfluous spin 1/2 carried by an interacting
spin-vector field. They did not find a satisfactory solution to this problem
\footnote{They specially consulted I.M. Gelfand on what such an unusual
symmetry could be \cite{Pcom}, but the great mathematician could not give them
any hint at that time (in the middle of the sixties).}. Now we know that this
mysterious gauge invariance is the local supersymmetry, while the corresponding
gauge theory is supergravity.

Ogievetsky quickly realized that supersymmetry is potentially capable of
providing an answer to his and Polubarinov's query about a self-consistent spin
3/2 theory. And it was he who initiated the study of this new type of symmetry
at LTP in the first half of the seventies. This paper is a brief (and
inevitably biased) account of the history of these studies for more than 30
years which passed since we became aware of supersymmetry, with focusing on the
milestones. Many of the results reviewed below  were paralleled and in some
cases rediscovered by other groups. Because of the lack of space and keeping in
mind a jubilee character of the present paper, I mainly cite the relevant works
of the Dubna group and frequently omit references to some important parallel
studies. I apologize for this incompleteness of the reference list.

\section{First studies: 1974 - 1980}

\noindent{\bf 2.1 Superspace: what it is and how it helps.} Any symmetry
implies some framework within which it admits a concise and instructive
realization. For instance, Poincar\'e symmetry can be naturally realized on
Minkowski space and fields given in it. For supersymmetry, such a natural
framework is superspace, an extension of some bosonic space by anticommuting
fermionic (Grassmann) coordinates. For the $N=1$ Poincar\'e supersymmetry \p{1}
it was actually introduced in one of the pioneering papers, \cite{VA}, as a
coset of the $N=1$ Poincar\'e supergroup over its bosonic Lorentz subgroup.
However, the fermionic coset parameters, in the spirit of the nonlinear
realizations method, were treated in \cite{VA} as Nambu-Goldstone fields
``living'' on Minkowski space. The treatment of the fermionic coordinates on
equal footing with $x^m$ as {\it independent} coordinates was suggested by
Salam and Strathdee \cite{SS} who considered fields on such an extended space
and showed that these fields naturally encompass the irreducible multiplets of
$N=1$ supersymmetry ($N=1$ supermultiplets). They named this space {\it
superspace} and fields on it {\it superfields}.

In $N=1$ superspace \be (x^m, \theta^\alpha, \bar\theta^{\dot \alpha}) \ee
$N=1$ supersymmetry \p{1} acts as shifts of Grassmann coordinates \be
\theta^\alpha{}' = \theta^\alpha + \epsilon^\alpha\,,  \quad
\bar\theta^{\dot\alpha}{}' = \bar\theta^{\dot\alpha} +
\bar\epsilon^{\dot\alpha}\,, \quad x^m{}' = x^m + i(\theta \sigma^m\bar\epsilon
- \epsilon\sigma^m\bar\theta) \lb{Tran} \ee where $\epsilon^\alpha,
\bar\epsilon^{\dot\alpha}$ are the mutually conjugated Grassmann transformation
parameters associated with the generators $Q_\alpha$ and $\bar Q_{\dot\alpha}$.
It is easy to check that the Lie bracket of two such transformations of $x^m$
yields a constant shift of $x^m$, in accord with relation \p{1}. A general
$N=1$ superfield is an unconstrained function $\Phi(x, \theta, \bar\theta)$
which transforms as
\bea \delta \Phi(x,\theta, \bar\theta) &=& -\delta
\theta^\alpha \frac{\partial}{\partial \theta^\alpha}\Phi(x,\theta,\bar\theta)
- \delta \bar\theta^{\dot\alpha}\frac{\partial}{\partial
\bar\theta^{\dot\alpha}}\Phi(x,\theta,\bar\theta)
- \delta x^m \frac{\partial}{\partial x^m}\Phi(x,\theta, \bar\theta) \nn \\
&& \equiv  i\left(\epsilon^{\alpha} Q_\alpha + \bar\epsilon_{\dot\alpha}
\bar Q^{\dot\alpha} \right)\Phi(x,\theta, \bar\theta)\,. \lb{Gener}
\eea
The generators $Q_\alpha, \bar Q_{\dot\alpha}$ can be checked to satisfy the anticommutation relation \p{1}.

The crucial feature of superfields is that they concisely encompass finite-component off-shell
field multiplets of the given supersymmetry.
As discovered by Salam and Strathdee, this key property is related
to the fact that $\theta^\alpha$ and $\bar\theta^{\dot\alpha}$ are anticommuting variables:
\be
\{\theta^\alpha, \theta^\beta \} =
\{\bar\theta^{\dot\alpha}, \bar\theta^{\dot\beta} \} = \{\theta^\alpha, \bar\theta^{\dot\beta} \} = 0\,.
\ee
These relations imply, in particular,
\be
(\theta^1)^2 = (\theta^2)^2 = 0 \quad (\mbox{and c.c.})\,.
\ee
Then, expanding $\Phi(x, \theta, \bar\theta)$ in a series over all possible monomials constructed from
$\theta^\alpha$ and $\bar\theta^{\dot\alpha}$, one observes that this series terminates at
the monomial $\theta^1\theta^2 \bar\theta^{\dot 1}\bar\theta^{\dot 2} \sim (\theta)^2 (\bar \theta)^2$,
where $(\theta)^2 = \epsilon_{\alpha\beta}\theta^\alpha\theta^\beta\,, (\bar\theta)^2
= \overline{(\theta)^2}\,$\footnote{We use the standard two-dimensional spinor notation,
with $\epsilon_{\alpha\beta} = -\epsilon_{\beta\alpha}, \epsilon_{12} = 1,
\epsilon^{\alpha\beta}\epsilon_{\beta\gamma} = \delta^\alpha_\gamma$ (and the same for dotted indices).}.
As a result, $\Phi(x,\theta, \bar\theta)$ contains $(8 + 8)$ fields: 8 bosonic fields and
8 fermionic fields:
\bea
\Phi(x,\theta, \bar\theta) &=& \phi(x) + \theta^\alpha \chi_\alpha(x) + \bar\theta_{\dot\alpha}\bar\chi^{\dot\alpha}(x)
+ \theta\sigma^n\bar\theta\,A_n(x) \nn \\
&& + \,(\bar\theta)^2\theta^\alpha \omega_\alpha(x) +
(\theta)^2\bar\theta_{\dot\alpha}\bar\omega^{\dot\alpha}(x) +
(\theta)^2(\bar\theta)^2 D(x)\,. \lb{Dcomp} \eea The precise transformation
laws of the component fields can be easily deduced from \p{Gener}. These fields
still form a {\it reducible} representation of $N=1$ supersymmetry. To make
$\Phi(x, \theta, \bar\theta)$ carry an irreducible supermultiplet, one needs to
impose on this superfield proper constraints covariant under $N=1$
supersymmetry. These constraints involve the covariant spinor derivatives, \be
D_\alpha =\frac{\partial}{\partial \theta^\alpha} +
i(\sigma^m\bar\theta)_\alpha \partial_m\,, \; \bar D_{\dot\alpha} =
-\frac{\partial}{\partial \bar\theta^{\dot\alpha}}-
i(\theta\sigma^m)_{\dot\alpha} \partial_m\,,\; \{D_\alpha, \bar
D_{\dot\alpha}\}= -2i (\sigma^m)_{\alpha\dot\alpha}\partial_m\,.  \lb{Der} \ee
These operators anticommute with the generators $Q_\alpha, \bar
Q_{\dot\alpha}$, so the result of their action on $\Phi(x, \theta, \bar\theta)$
is again a superfield. The covariant constraints singling out two irreducible
multiplets contained in a general unconstrained $\Phi(x, \theta, \bar\theta)$
are as follows \bea (\mbox{a})\, \bar D_{\dot\alpha} \Phi_{(1)}(x, \theta,
\bar\theta) = 0\,,(\mbox{or} \, D_{\alpha} \bar\Phi_{(1)}= 0 )\; (\mbox{b}) \,
(D)^2 \Phi_{(2)}(x,\theta,\bar\theta) = (\bar D)^2 \Phi_{(2)}(x, \theta,
\bar\theta) = 0\,. \lb{ConstR} \eea Using the appropriate projection operators,
the general real superfield $\Phi(x,\theta, \bar\theta)$ can be decomposed into
the irreducible pieces as follows:
\be \Phi = \Phi_{(1)} + \bar\Phi_{(1)} +
\Phi_{(2)}\,. \lb{DeC} \ee

This decomposition is an analog of the well-known decomposition of $4D$ vector
field into the longitudinal and transversal parts (spins 0 and 1). In the case
of supersymmetry, the notion of spin is generalized to the {\it superspin}. The
constrained superfields $\Phi_{(1)}$ and $\Phi_{(2)}$ can be shown to possess
definite superspins, 0 and 1/2, respectively. The superfield constraint
(\ref{ConstR}a) admits a nice geometric solution. Namely, making the complex
change of the superspace coordinates \be (x^m, \theta^\alpha,
\bar\theta^{\dot\alpha}) \quad \Rightarrow \quad (x^m_L = x^m +
i\theta\sigma^m\bar\theta\,, \theta^\alpha\,, \bar\theta^{\dot\alpha})\,, \ee
one finds that $\bar D_{\dot\alpha}$ is ``short'' in this new (``left-chiral'')
basis \be \bar D_{\dot\alpha} = -\frac{\partial}{\partial
\bar\theta^{\dot\alpha}}\,, \ee and (\ref{ConstR}a) becomes the Grassmann
Cauchy-Riemann condition stating that $\Phi_{(1)}$ is independent of the half
of Grassmann coordinates in this basis: \be \frac{\partial}{\partial
\bar\theta^{\dot\alpha}} \Phi_{(1)}(x, \theta, \bar\theta)=0 \quad \Rightarrow
\quad \Phi_{(1)}(x, \theta, \bar\theta) = \varphi(x^m_L, \theta^\alpha)\,. \ee
It is easy to directly check that the set $(x^m_L, \theta^\alpha)$ is closed
under the supertranslations \p{Tran} and so forms a complex invariant space of
the $N=1$ Poincar\'e supergroup, chiral superspace. The $\theta$ expansion of
the superfield $\varphi(x_L,\theta)$, {\it chiral} $N=1$ superfield
\cite{Ferr}, directly yields the scalar $N=1$ supermultiplet of fields: \be
\varphi(x^m_L, \theta^\alpha) = \varphi(x_L) + \theta^\alpha \psi_\alpha(x_L) +
(\theta)^2 F(x_L)\,, \label{AuxF} \ee where $\varphi(x_L)$ and $F(x_L)$ are two
complex scalar fields and $\psi_\alpha(x_L)$ is a two-component left-chiral
Weyl spinor.

The basic advantages of using off-shell superfields are as follows.

First of all, their SUSY transformation laws do not depend on the dynamics,
i.e. are the same whatever the invariant action of the involved fields is. An
important property of superfields is the presence of the so-called auxiliary
fields in their $\theta$ expansion, which is necessary for the off-shell
closure of the SUSY algebra on the component fields. In the example \p{AuxF} it
is just the field $F(x_L)$. Ascribing the canonical dimensions 1 and $3/2$ to
the ``physical fields'' $\varphi$ and $\psi_\alpha$ and taking into account
that $[\theta] = -1/2$, one finds that $[F] = 2$, whence it follows that $F$
should enter any $D=4$ action without derivatives. In other words, its equation
of motion is always algebraic and serves to express $F$ in terms of the
physical fields (or to put $F$ equal to a constant or zero). Since SUSY mixes
this algebraic equation with those for physical fields, it closes on the
physical fields only modulo their equations of motion. As a result, the
realization of SUSY on the physical fields depends on the choice of the
invariant action, and for this reason it proves very difficult to construct
invariant actions with making use of the physical fields only.

On the other hand, any product of superfields, with or without $x$- or spinor
derivatives on them, is again a superfield. The second crucial property of
off-shell superfields is that the component field appearing as a coefficient of
the highest-degree $\theta$ monomial always transforms as a total
$x$-derivative of the lower-order component fields. Hence, its integral over
Minkowski space is SUSY {\it invariant} and so is a candidate for an invariant
action. Forming products of some basic elementary superfields and using the
property that these products are superfields on their own, one can be sure that
the (composite) component fields appearing as coefficients of the highest-order
$\theta$ monomials in these products are transformed by a total derivative. So
the invariant actions can be constructed as Minkowsky space integrals of these
composite fields. In other words, the superfield approach provides a universal
way of searching for supersymmetric actions.

The remarkable features of the superfield approach listed above led V.I.
Ogievetsky to rapidly realize how indispensable it promises to be for exploring
geometric and quantum properties of supersymmetric theories. In the middle of
the seventies, he started to actively work on the superspace methods, together
with his disciples Luca Mezincescu from Bucharest and Emery Sokatchev
from Sofia. \vspace{0.3cm}

\noindent{\bf 2.2 Action principle in superspace.} In \cite{OM1} Ogievetsky and
Mezincescu proposed an elegant way of writing down the invariant superfield
actions. As mentioned above, the invariant actions can be constructed as the
$x$-integrals of the coefficients of the highest-degree $\theta$ monomials in
the appropriate products of the involved superfields. The question was how to
extract these components in a manifestly supersymmetric way. Ogievetsky and
Mezincescu proposed to use the important notion of Berezin integral \cite{Ber}
for this purpose. In fact, Berezin integration is equivalent to the Grassmann
differentiation and, in the case of $N=1$ superspace, is defined by the rules
\be \int d\theta_\alpha\, \theta^\beta = \delta^\beta_\alpha\,, \quad \int
d\theta^\alpha\, 1 = 0\,, \quad \{d\theta_\alpha, d\theta_\beta \}
=\{\theta_\alpha, d\theta_\beta \} = 0\,.\lb{BerInt} \ee
It is easy to see
that, up to the appropriate normalization, \be \int d^2 \theta \,(\theta)^2 =
1\,, \;\; \int d^2 \bar\theta\, (\bar\theta)^2 = 1\,, \;\; \int d^2\theta
d^2\bar\theta\, (\theta)^4 = 1\,, \lb{BerInt2} \ee and, hence, Berezin
integration provides the efficient and manifestly supersymmetric way of
singling out the coefficients of the highest-order $\theta$ monomials. For
example, the simplest invariant action of chiral superfields can be written as

\be S \sim \int d^4x d^4\theta\, \varphi(x_L, \theta)\bar\varphi (x_R,
\bar\theta)\,, \quad x^m_R = \overline{(x_L^m)} = x^m - i\theta \sigma^m
\bar\theta\,. \lb{ChirAct} \ee

Using \p{AuxF} and \p{BerInt2}, it is easy to
integrate over $\theta, \bar\theta$ in \p{ChirAct} and, discarding total
$x$-derivatives,  to obtain the component form of the action \be S \sim \int
d^4 x \left(\partial^m\bar\varphi \partial_m\varphi -\sfrac{i}{2}
\psi\sigma^m\partial_m\bar\psi + F\bar F \right).\lb{FreeWZ} \ee It is just the
free action of the massless scalar $N=1$ multiplet. One can easily generalize
it to the case with interaction by choosing the Lagrangian as an arbitrary
function $K(\bar\varphi, \varphi)$ and adding independent potential terms \be
\sim \int d^4x_L d^2\theta\, P(\varphi) + \mbox{c.c.}\,, \lb{PoT} \ee which in
components produce mass terms, scalar potentials, and fermionic Yukawa coupling
for the physical fields after elimination of the auxiliary fields $F, \bar F$
in a sum of the superfield kinetic and potential terms. The sum of \p{ChirAct}
and the superpotential term \p{PoT} with $P(\varphi) \sim g\varphi^3 + m
\varphi^2$ corresponds to the Wess-Zumino model \cite{WZu} which was the first
example of nontrivial $N=1$ supersymmetric model and the only renormalizable
model of scalar $N=1$ multiplet. Ogievetsky and Mezincescu  argued in
\cite{OM1} that the representation of the action of the Wess-Zumino model in
terms of Berezin integral was very useful and suggestive while developing the
superfield perturbation theory for it \footnote{One can show that all quantum
corrections have the form of the integral over the whole $N=1$ superspace, so
the superpotential term (and, hence, the parameters $g$ and $m$) is not
renormalized. This statement is the simplest example of the so-called
non-renormalization theorems.}.

In 1975, Ogievetsky and Mezincescu wrote a comprehensive review on the basics of supersymmetry and
superspace approach \cite{Rev}. Until present it remains one of the best introductory reviews in the field.
\vspace{0.3cm}

\noindent{\bf 2.3 Superfields with higher superspins and new supergauge
theories.} The next benchmark became Sokatchev's work \cite{ES1} where the
general classification of $N=1$ superfields with respect to superspin was
given, and the corresponding irreducibility superfield constraints
(generalizing \p{ConstR}) together with the relevant projection operators on
definite superspins were given in an explicit form. In the pioneering paper
\cite{SS}, the decomposition into the superspin-irreducible parts was discussed
in detail only for a scalar $N=1$ superfield. Higher superspins are carried by
superfields with external Lorentz indices. Like in the case of bosonic gauge
theories, the requirement of preserving definite superspins by interacting
superfields was expected to fully determine the structure of the corresponding
action and the gauge group intended to make harmless extra superspins carried
by the given off-shell superfield. In fulfilling this program of research, the
formalism of the projection operators of \cite{ES1} proved to be indispensable.

An $N=1$ superextension of the Yang-Mills theory was constructed in \cite{SYM}.
It was shown that the fundamental object (prepotential) carrying the
irreducible field content of the off-shell $N=1$ vector multiplet (gauge field
$b_m(x)$, gaugino $\psi_\alpha (x), \bar\psi_{\dot\alpha}(x)$ and the auxiliary
field $D(x)$, all taking values in the adjoint representation of gauge group)
is the real scalar superfield $V(x,\theta, \bar\theta)$ with certain gauge
freedom. The latter, in the abelian case, is given by the transformations \be
V'(x,\theta, \bar\theta) = V(x,\theta, \bar\theta) +
\sfrac{i}{2}\left(\Lambda(x_L, \theta) - \bar\Lambda(x_R, \bar\theta\right)\,,
\lb{GaugeV} \ee where $\Lambda$ and $\bar\Lambda$ are mutually conjugated
superfield parameters ``living'' as unconstrained functions on the left and
right $N=1$ chiral subspaces. Any component in $V(x,\theta, \bar\theta)$ which
undergoes an additive shift by a gauge parameter, can be fully removed by
fixing this parameter; proceeding in this way, one can show that the maximally
reduced form of $V(x,\theta, \bar\theta)$ (Wess-Zumino gauge) is  as follows
\bea && V(x,\theta, \bar\theta) = \theta\sigma^n\bar\theta\,A_n +
(\bar\theta)^2\theta^\alpha \psi_\alpha +
(\theta)^2\bar\theta_{\dot\alpha}\bar\psi^{\dot\alpha}
+ (\theta)^2(\bar\theta)^2 D\,, \quad \delta A_n = \partial_n \lambda_0\,, \lb{WZ} \\
&&\lambda_0 \equiv
-\sfrac{1}{2}(\Lambda + \bar\Lambda)|_{\theta = \bar\theta = 0}\,. \nonumber
\eea
The fields in \p{WZ} are recognized as the irreducible off-shell $N=1$ vector multiplet (superspin 1/2).

Ogievetsky and Sokatchev asked whether there exist more complicated superfield
gauge theories, with the prepotentials having extra Lorentz indices and so
carrying other definite superspins in interaction. Using the formalism of the
projection operators developed in \cite{ES1}, they firstly tried to construct a
self-contained theory of spinor gauge superfield $\Psi_\alpha (x,\theta,
\bar\theta), \bar\Psi_{\dot\alpha} (x,\theta, \bar\theta)$ \cite{ES2} as an
alternative to the standard $N=1$ gauge theory, with the gauge vector being in
the same irreducible multiplet with a massless spin 3/2 field. They constructed
a self-consistent free action for such a spinor superfield, but failed to
promote some important gauge symmetry of it to a non-Abelian interacting case.
The reason for this failure was realized later on: a self-consistent theory of
interacting massless Rarita-Schwinger field should be supergravity which
necessarily includes Einstein gravity as a subsector.

Searching for a self-consistent theory of massless vector superfield (carrying superspins 3/2 and 1/2)
turned out to be more suggestive. This superfield $H^{n}(x,\theta, \bar\theta)$ encompasses, in its
component field expansion, massless tensor field $e_{a}^n$ and spin-vector field $\psi_{\alpha}^n\,$,
$$
H^n = \theta\sigma^a\bar\theta e_{a}^n + (\bar\theta)^2 \theta^\alpha \psi_{\alpha}^n +
(\theta)^2 \bar\theta_{\dot\alpha}\bar\psi^{\dot\alpha n} + \ldots\,,
$$
which could naturally be identified with the graviton and gravitino fields. In
\cite{ES2} Ogievetsky and Sokatchev put forward the hypothesis that the correct
``minimal'' $N=1$ superfield supergravity should be a theory of gauge
axial-vector superfield $H^m(x,\theta, \bar\theta)$ generated  by the conserved
supercurrent. The latter unifies into an irreducible $N=1$ supermultiplet the
energy-momentum tensor and spin-vector current associated with the
supertranslations (see \cite{SCurr}, \cite{SEM} and refs. therein). Ogievetsky
and Sokatchev relied upon the clear analogy with the Einstein gravity which can
be viewed as a theory of massless tensor field generated by the conserved
energy-momentum tensor. The whole Einstein action and its non-Abelian $4D$
diffeomorphism gauge symmetry can be uniquely restored step-by-step, starting
with a free action of symmetric tensor field and requiring its source
(constructed from this field and its derivatives, as well as from matter
fields) to be conserved \cite{OP2}. In \cite{ES3} this Noether procedure was
applied to the free action of $H^m(x,\theta, \bar\theta)\,$. The first-order
coupling of $H^m$ to the conserved supercurrent of the matter chiral superfield
was restored and superfield gauge symmetry generalizing bosonic diffeomorphism
symmetry was identified at the linearized level. The geometric meaning of this
supergauge symmetry and its full non-Abelian form were revealed by Ogievetsky
and Sokatchev later, in the remarkable papers \cite{ESsg,ESsg1}. Before
dwelling on this, let me mention a few important parallel investigations on
$N=1$ SUSY performed in our Sector approximately at the same time, i.e. in the
second half of the seventies and beginning of the eighties. \vspace{0.3cm}

\noindent{\bf 2.4 General relation between linear and nonlinear realizations of $N=1$
SUSY.} One of the first known realizations of $N=1$ SUSY was its nonlinear
(Volkov-Akulov) realization \cite{VA}
\be
y^m{\,}' = y^m + i[\lambda(y) \sigma^m\bar\epsilon - \epsilon\sigma^m\bar\lambda(y)]\,,
\quad \lambda^\alpha{\,}'(y{\,}') =
\lambda^\alpha(y) + \epsilon^\alpha\,, \;\bar\lambda^{\dot\alpha}{\,}'(y{\,}') =
\bar\lambda^{\dot\alpha}(y) + \bar\epsilon^{\dot\alpha}\,, \label{NL}
\ee
where the corresponding Minkowski space coordinate is denoted by $y^m$ to distinguish it from $x^m$
corresponding to the superspace realization \p{Tran}. The main difference between \p{NL} and \p{Tran} is that \p{NL}
involves the $N=1$ Goldstone fermion (goldstino) $\lambda (y)$ the characteristic feature of which
is the inhomogeneous transformation law under supertranslations, which corresponds to the spontaneously
broken SUSY. It is a field given on Minkowski space, while $\theta^\alpha$ in \p{Tran} is an independent
Grassmann coordinate, and $N=1$ superfields support a linear realization of $N=1$ SUSY.
The invariant action of $\lambda, \bar\lambda$ is \cite{VA}:
\be
S_{(\lambda)} = \sfrac{1}{f^2}\int d^4 y \, \det E^a_m\,, \quad E^a_m = \delta^a_m +
i\left(\lambda\sigma^a\partial_m\bar\lambda - \partial_m\lambda\sigma^a\bar\lambda \right).\label{VAact}
\ee
where $f$ is a coupling constant ($[f] = -2\,$).

The natural question was what is the precise relation between the nonlinear and
superfield (linear) realizations of the same $N=1$ SUSY. We with my friend and
co-worker Sasha Kapustnikov (now late) were the first to pose this question and
present the explicit answer \cite{IKa1}-\cite{IKa3}. We showed that, given the
Goldstone fermion $\lambda (y)$ with the transformation properties \p{NL}, the
relation between two types of the $N=1$ SUSY realizations, \p{Tran} and \p{NL},
is given by the following invertible change of the superspace coordinates: \be
x^m = y^m + i\left[\theta\sigma^m\bar\lambda(y) -
\lambda(y)\sigma^m\bar\theta\right], \quad \theta^\alpha =
\tilde{\theta}^\alpha + \lambda^\alpha(y)\,, \; \bar\theta^{\dot\alpha} =
\tilde{\bar\theta}^{\dot\alpha} + \bar\lambda^{\dot\alpha}(y)\,, \lb{NLchang}
\ee where \be \tilde{\theta}^\alpha{\,}' = \tilde{\theta}^\alpha\,. \lb{Inert}
\ee Then the transformations \p{NL} imply for $(x^m, \theta^\alpha,
\bar\theta^{\dot\alpha})$ just the transformations \p{Tran} and, vice-versa,
\p{Tran} imply \p{NL}. Using \p{NLchang}, any linearly transforming superfield
can be put in the new ``splitting'' basis \be \Phi(x, \theta, \bar\theta) =
\tilde{\Phi}(y, \tilde{\theta}, \tilde{\bar\theta})\,.\lb{SupSpl} \ee Since
$\tilde{\theta}^\alpha$ is ``inert'' under $N=1$ SUSY, eq. \p{Inert}, the
components of $\tilde{\Phi}$ transform as ``sigma-fields'' \be \delta \phi(y) =
-i[\lambda(y) \sigma^m\bar\epsilon -
\epsilon\sigma^m\bar\lambda(y)]\partial_m\phi(y)\,, \quad \mbox{etc}\,, \ee
independently of each other, whence the term ``splitting'' for this basis. As
demonstrated in \cite{IKa3}, irrespective of the precise mechanism of
generating goldstino in a theory with the linear realization of spontaneously
broken $N=1$ SUSY, the corresponding superfield action can be rewritten in the
splitting basis (after performing integration over the inert Grassmann
variables) as \be S_{lin} \sim \int d^4 y \det E^a_m \left [ 1 + {\cal
L}(\sigma, \nabla_a\sigma, ...)\right]. \ee Here ${\cal L}$ is a function of
the ``sigma'' fields and their covariant derivatives $\nabla_a =
E^m_a\partial_m\,$ only, while $\lambda^\alpha(y)$ is related to the goldstino
of the linear realization through a field redefinition. Thus, the Goldstone
fermion is always described by the universal action \p{VAact}, independently of
details of the given dynamical theory with the spontaneous breaking of $N=1$
supersymmetry, in the spirit of the general theory of nonlinear realizations.

The transformation \p{NLchang}, \p{SupSpl} can be easily generalized to chiral superfields
and to higher $N$. It proved very useful for exhibiting the low-energy structure of
theories with spontaneously broken SUSY and in some other problems.
It was generalized to the case of local $N=1$ SUSY in \cite{IKa4}.
\vspace{0.3cm}

\noindent{\bf 2.5 AdS$_4$ superspace.} Soon after the $N=1$ Poincar\'e
supersymmetry was discovered, there was found $N=1$ superextension of another
important $D=4$ group, conformal group $SO(2,4) \sim SU(2,2)\,$. The latter was
known to play an important role in quantum field theory (specifying the
structure of Green functions in some massless $D=4$ models), as well as in
gravity which, e.g.,  can be regarded as a theory following from the
spontaneous breaking of the local conformal group with the Goldstone dilaton
field as a ``compensator''  (see e.g. \cite{FT}). This was the main motivation
for considering $N=1$ superconformal group $SU(2,2|1)$ (and its higher $N$
analogs $SU(2,2|N)$). Later on, the gauge versions of these symmetries were
used to construct extended supergravities.

An important property of the conformal group is that it admits a natural action
in the conformally-flat $D=4$ space-times, with the distances  related to the
Minkowski interval by a Weyl factor. The corresponding groups of motion are
subgroups of the conformal group. This class of spaces includes anti-de Sitter
and de Sitter spaces AdS$_4 \sim \frac{SO(2,3)}{SO(1,3)}$ and dS$_4 \sim
\frac{SO(1,4)}{SO(1,3)}\,$. One could expect that the property of conformal
flatness is generalized to superspaces. While the dS$_4$ spinor comprises 8
independent components, no such doubling as compared to the Minkowski space
occurs for AdS$_4$: the AdS$_4$ spinor is the Weyl one with two complex
components. Keeping this in mind, the corresponding SUSY was expected to be
similar to \p{1}. There was an urgent necessity to construct a self-consistent
superfield formalism for AdS$_4$ SUSY, and in 1978 we turned to this problem
with my PhD student A. Sorin from the Dniepropetrovsk State University
\footnote{Now - Deputy Director of BLTP Prof. A.S. Sorin.}.

$N=1$ AdS$_4$ superalgebra is $osp(1|4) \subset su(2,2|1)$, and it is  defined
by the following (anti)commutation relations:
\bea && \{Q_\alpha, \bar
Q_{\dot\alpha}\} = 2(\sigma^m)_{\alpha\dot\alpha}P_m\,, \quad
\{Q_\alpha, Q_{\beta}\} = m (\sigma^{mn})_{\alpha\beta}L_{[m,n]}\,, \nn \\
&& [ Q_\alpha, P_m] = \sfrac{m}{2}(\sigma_m)_{\alpha\dot\alpha}\bar Q^{\dot\alpha}\,, \quad
[P_m, P_n] = -i m^2 L_{[m,n]}\,. \label{ADS4}
\eea
Here $(\sigma^{mn})_\alpha^\beta = \sfrac{i}{2}(\sigma^m\tilde{\sigma}^n
- \sigma^n\tilde{\sigma}^m)_\alpha^\beta\,$,
$(\tilde{\sigma}^m)^{\dot\alpha\beta}
= \epsilon^{\dot\alpha\dot\omega}\epsilon^{\beta\gamma}(\sigma^m)_{\gamma\dot\omega}$,
$m\sim r^{-1}$ is the inverse radius of AdS$_4\,$ and $L_{[m,n]}$
are generators of the Lorentz $SO(1,3)$ subgroup of $SO(2,3)\propto (P_m, L_{[m,n]})\,$. To eqs. \p{ADS4}
one should add complex-conjugate relations and (trivial) commutators with $L_{[m,n]}\,$.
In the limit $m \rightarrow 0$ ($r \rightarrow \infty$), \p{ADS4} go over into  \p{1}.

In \cite{IS1,IS2}, for constructing $OSp(1|4)$ covariant superfield formalism
we applied a powerful method of Cartan forms (viz. the coset method) which
allowed us to find the true AdS$_4$ analogs of the general and chiral $N=1$
superfields, as well as the vector and spinor covariant derivatives, invariant
superspace integration measures, etc. Having developed the AdS$_4$ superfield
techniques, we constructed the $OSp(1|4)$ invariant actions generalizing the
actions of the Wess-Zumino model and $N=1$ SYM theory. Just to give a feeling
what such actions look like, I present here an analog of the free massless
action \p{FreeWZ} of $N=1$ scalar multiplet, with the auxiliary fields
eliminated by their equations of motion: \be S \sim \int d^4 x\,
a^4(x)\left(\partial^m\bar\varphi \partial_m\varphi -\sfrac{i}{4}
\psi\sigma^m\nabla_m\bar\psi +\sfrac{i}{4} \nabla_m\psi\sigma^m\bar\psi  +
2m^2\,\varphi\bar\varphi  \right). \lb{ADSWZ} \ee Here $a(x) = \sfrac{2}{1 +
m^2 x^2}$ is a scalar factor specifying the AdS$_4$ metric in a
conformally-flat parametrization, $ds^2 = a^2(x)\eta_{mn} dx^m d x^n\,$, and
$\nabla_m = a^{-1}\partial_m\,$. Taking into account that $m^2 =
-\sfrac{1}{12}R$ where $R$ is the scalar curvature of AdS$_4$, this action is
the standard form of the massless scalar field action in a curved background.

In \cite{IS2} we thoroughly studied the vacuum structure of the general massive
AdS$_4$ Wess-Zumino model, which turned out to be much richer as compared to
the standard ``flat'' Wess-Zumino model due to the presence of the``intrinsic''
mass parameter $m\,$. We also showed that both the AdS$_4$ massless Wess-Zumino
model and super YM model can be reduced to their flat $N=1$ super Minkowski
analogs via some superfield transformation generalizing the Weyl transformation
\be \varphi(x)  = a^{-1}(x) \tilde\varphi(x), \quad \psi^\alpha(x) =
a^{-3/2}(x)\tilde{\psi}^\alpha(x)\,,\lb{Weyl} \ee which reduces \p{ADSWZ} to
\p{FreeWZ}. The existence of the superfield Weyl transformation was an
indication of the superconformal flatness of the AdS$_4$ superspace (although
this property has been proven only recently \cite{SFlat}).

Afterwards, the simplest supermultiplets of $OSp(1|4)$ derived for the first
time in \cite{IS1} from the superfield formalism and the corresponding
projection operators were used, e.g., in \cite{WS} to give an algebraic meaning
to the superfield constraints of $N=1$ supergravity. The interest in $OSp(1|4)$
supersymmetry and the relevant model-building has especially grown up in recent
years in connection with the famous Maldacena's AdS/CFT conjecture.

\section{Complex geometry of $N=1$ supergravity}
Poincar\'e $N=1$ supergravity (SG) as a theory of interacting gauge vierbein
field $e_m^a(x) = \delta^a_m + \kappa h^a_m(x)$ (graviton, with $\kappa$ being
Einstein constant) and spin-vector field $\psi^\mu_m(x),
\bar\psi^{\dot\mu}_m(x)$ (gravitino) and possessing, in addition to D=4
diffeomorphisms, also a local supersymmetry, was discovered in \cite{SG}. It
was an urgent problem to find a full off-shell formulation of $N=1$ SG, i.e.,
to complete the set of physical fields $e, \psi$ to an off-shell multiplet by
adding the appropriate auxiliary fields and/or to formulate $N=1$ SG in
superspace, making all its symmetries manifest.

One of the approaches to $N=1$ SG in superspace was based on considering the
most general differential geometry in $N=1$ superspace. One defines supervielbeins, supercurvatures
and supertorsions which are covariant under arbitrary $N=1$ superdiffeomorphisms, and then imposes
the appropriate constraints, so as to end up with the minimal set of off-shell $N=1$ superfields
encompassing the irreducible field content of SG \cite{W}. Another approach is
to reveal the fundamental minimal gauge group of SG and the basic unconstrained SG
prepotential, an analog of $N=1$ SYM prepotential \p{GaugeV}. This was just the strategy which Ogievetsky
and Sokatchev kept to in \cite{ESsg} to discover a beautiful geometric formulation of
the conformal and ``minimal'' Einstein $N=1$ SG.

It is based on the generalization of the notion of flat $N=1$ chirality to the
curved case. The flat chiral $N=1$ superspace $(x_L^m, \theta^\mu_L)$ possesses
the complex dimension $(4|2)$ and includes the $N=1$ superspace $(x^m,
\theta^\mu, \bar\theta^{\dot\mu})$ as a real $(4|4)$ dimensional hypersurface
defined by the following embedding conditions \bea \mbox{(a)} \;\; x^m_L +
x^m_R = 2 x^m\,, \quad \mbox{(b)}\;\;  x^m_L - x^m_R =
2i\theta\sigma^m\bar\theta\,, \quad \theta^\mu_L = \theta^\mu\,, \;
\bar\theta^{\dot\mu}_R = \bar\theta^{\dot\mu}\,, \label{Emb} \eea and $x^m_R =
\overline{(x^m_L)}, \bar\theta^{\dot\mu}_R = \overline{(\theta_L^\mu)}\,$. It
turned out that the underlying gauge group of conformal $N=1$ SG is just the
group of general diffeomorphisms of the chiral superspace: \be \delta x^m_L =
\lambda^m(x_L, \theta_L)\,, \quad \delta \theta^\mu_L = \lambda^\mu(x_L,
\theta_L)\,,\lb{ChirDif} \ee with $\lambda^m, \lambda^\mu$ being arbitrary
complex functions of their arguments. The fermionic part of the embedding
conditions \p{Emb} remains unchanged while the bosonic one is generalized to
\be \mbox{(a)} \;\;x^m_L + x^m_R = 2 x^m\,, \quad \mbox{(b)}\;\;  x^m_L - x^m_R
= 2i H^m(x, \theta, \bar\theta)\,. \lb{Emb1} \ee The basic gauge prepotential
of conformal $N=1$ SG is just the axial-vector superfield $H^m(x,\theta,
\bar\theta)$ in \p{Emb1}. It specifies the superembedding of real $N=1$
superspace as a hypersurface into the complex chiral $N=1$ superspace $(x_L^m,
\theta^\mu_L)\,$ and so possesses a nice geometric meaning.  Through relations
\p{Emb1}, the transformations \p{ChirDif} generate field-dependent nonlinear
transformations of the $N=1$ superspace coordinates $(x^m, \theta^\mu,
\bar\theta^{\dot\mu})$ and of the superfield $H^m(x, \theta, \bar\theta)\,$.
The field content of $H^m$ can be revealed in the WZ gauge which requires
knowing only the linearized form of the transformations: \bea \delta^* H^m &=&
\sfrac{1}{2i}\left[\lambda^m(x + i\theta\sigma\bar\theta, \theta) -
\bar\lambda^m(x - i\theta\sigma\bar\theta, \bar\theta)\right] \nn \\
&& -\, \lambda(x + i\theta\sigma\bar\theta, \theta)\sigma^m\bar\theta -
\theta\sigma^m \bar\lambda(x - i\theta\sigma\bar\theta, \bar\theta)\,.\lb{WZH1}
\eea Here we took into account the presence of the ``flat'' part
$\theta\sigma^m\bar\theta$ in $H^m = \theta\sigma^a\bar\theta(\delta^m_a +
\kappa h^m_a) + \ldots \,$. An easy calculation yields the WZ gauge form of
$H^m$ as \be H^m_{WZ} = \theta\sigma^a\bar\theta\,e^m_a + (\bar\theta)^2
\theta^\mu \psi^m_\mu + (\theta)^2 \bar\theta_{\dot\mu}\bar\psi^{m\dot\mu} +
(\theta)^2(\bar\theta)^2 A^m\,. \label{WZH} \ee Here one finds the vierbein
$e^m_a$ presenting the conformal graviton (gauge-independent spin 2 off-shell),
the gravitino $\psi^m_\mu$ (spin $(3/2)^2$), and the gauge field $A^m$ (spin 1)
of the local $\gamma_5$ R-symmetry, just $(8 +8)$ off-shell degrees of freedom
forming the superspin 3/2 $N=1$ Weyl multiplet.

The Einstein $N=1$ SG can now be deduced in two basically equivalent ways.
The first one was used in the
original paper \cite{ESsg} and it is to restrict the group \p{ChirDif} by the constraint
\be
\partial_m\lambda^m - \partial_\mu \lambda^\mu = 0\,,\lb{Ber1}
\ee which is the infinitesimal form of the requirement that the integration
measure of chiral superspace $(x_L, \theta^\mu)$ is invariant. One can show
that, with this constraint, the WZ form of $H^m$ collects two extra scalar
auxiliary fields, while $A^m$ ceases to be gauge and also becomes an auxiliary
field. On top of this, there disappears one fermionic gauge invariance
(corresponding to conformal SUSY) and, as a result, spin-vector field starts to
carry  12 independent components. So one ends up with the (12 + 12) off-shell
multiplet of the so-called ``minimal'' Einstein SG \cite{AuxSG}.

Another, more suggestive way to come to the same off-shell content is to use
the compensator ideology which can be traced back to the interpretation of
Einstein gravity as conformal gravity with the compensating (Goldstone) scalar
field \cite{FT}. Since the group \p{ChirDif} preserves the chiral superspace,
in the local case one can still define a chiral superfield $\Phi(x_L, \theta)$
as an unconstrained function on this superspace and ascribe to it the following
transformation law \be \delta \Phi = -\sfrac{1}{3}\left(\partial_m\lambda^m -
\partial_\mu \lambda^\mu\right) \Phi\,,\lb{TranphiL} \ee where the specific
choice (-1/3) of the conformal weight of $\Phi$ is needed for constructing the
invariant SG action. Assuming that the vacuum expectation value of $\Phi$ is
non-vanishing and recalling the $\theta$ expansion \be \Phi = <f> + \tilde{f}+ i g +
\theta^\mu \chi_\mu + (\theta)^2(S+ iP)\,, \quad <f> \neq 0\,,  \lb{phi} \ee
one observes from the transformation law \p{TranphiL} that the fields $\tilde{f}, g$
and $\chi_\mu$ can be gauged away, thus fully ``compensating'' dilatations,
R-transformations and conformal supersymmetry. The fields $S$ and $P$ and the non-gauge field
$A^m$ coming from $H^m$ constitute the set of auxiliary fields. Together with other fields from the
appropriate WZ gauge for $H^m(x,\theta, \bar\theta)$ they yield
the required off-shell $(12 + 12)$ representation.

The basic advantage of the compensating method is that it allows one to easily write the action of the minimal
Einstein SG as an invariant action of the compensator $\Phi$ in the background of the Weyl multiplet carried
by $H^m$:
\bea
S_{SG} &=& - \sfrac{1}{\kappa^2}\int d^4x d^2\theta d^2\bar\theta\, E \,
\Phi(x_L, \theta)\bar\Phi(x_R, \bar\theta) \nn \\
&& + \,\xi \left(\int d^4x_L d^2\theta \Phi^3(x_L, \theta) + \mbox{c.c.} \right). \lb{SGact}
\eea
Here $E$ is a density constructed from $H^m$ and its derivatives \cite{ESsg1}, such that
its transformation cancels the total weight transformation of the integration measure
$d^4xd^2\theta d^2\bar\theta$ and the product of chiral compensators. In components,
the first term in \p{SGact} yields the minimal Einstein
$N=1$ SG action without cosmological term, while the second term in \p{SGact} is the superfield form of
the cosmological term $\sim \xi\,$.\footnote{The original Ogievetsky-Sokatchev differential geometry formalism
and invariant action \cite{ESsg1} amount to some specific gauge choice in \p{SGact}.}

Later on, many other off-shell component and superfield versions of $N=1$ SG were discovered.
They mainly differ in the choice of the compensating supermultiplet. This variety of compensating superfields
is related to the fact that the same on-shell scalar $N=1$ multiplet admits variant off-shell
representations.

The Ogievetsky-Sokatchev formulation of $N=1$ SG was one of the main
indications that the notion of chiral superfields and chiral superspace play
the key role in $N=1$ supersymmetry. Later it was found that the superfield
constraints of $N=1$ SG have the nice geometric meaning: they guarantee the
existence of chiral $N=1$ superfields in the curved case, once again pointing
out the fundamental role of chirality in $N=1$ theories. The constraints
defining the $N=1$ SYM theory can also be derived from requiring chiral
representations to exist in the full interaction case. The parameters of the
$N=1$ gauge group are chiral superfields (see \p{GaugeV}), so this group
manifestly preserves the chirality. The geometric meaning of $N=1$ SYM
prepotential $V(x,\theta, \bar\theta)$ was discovered in \cite{I11}. By analogy
with $H^n(x,\theta, \bar\theta)$, the superfield $V$ specifies a real $(4|4)$
dimensional hypersurface, this time in the product of $N=1$ chiral superspace
and the internal coset space $G^c/G$, where $G^c$ is complexification of the
gauge group $G\,$. At last, chiral superfields provide the most general
description of $N=1$ matter since any variant off-shell representation of $N=1$
scalar multiplet is related to chiral multiplet via duality transformation.

Soon after revealing the nice geometric formulation of $N=1$ SG described
above, there arose a question as to how it can be generalized to the most
interesting case of extended supergravities and, first of all, to $N=2$
supergravity. To answer this question, it proved necessary to understand what
the correct generalization of $N=1$ chirality to $N\geq 2$ SUSY is and to
invent a new sort of superspaces, the harmonic ones.

\section{Extended SUSY and harmonic superspace}

\noindent{\bf 4.1 Difficulties.} The basic problem with extended superspace
$(x^m, \theta_{i}^\alpha, \bar\theta^{\dot\alpha i})$ was that the
corresponding superfields, due to a large number of Grassmann coordinates,
contain too many irreducible supermultiplets. So they should be either strongly
constrained or subjected to some powerful gauge groups, with \'a priori unclear
geometric meaning. Another problem was that some constraints imply the
equations of motion for the involved fields before assuming any invariant
action for them. For instance, in the $N=2$ case ($i=1,2$) the simplest matter
multiplet (analog of $N=1$ chiral multiplet) is the hypermultiplet which is
represented by a complex $SU(2)$ doublet superfield $q^{i}(x, \theta,
\bar\theta)$ subjected to the constraints \be D^{(i}_\alpha q^{k)} = \bar
D^{(i}_{\dot\alpha} q^{k)} = 0\,. \lb{HypC} \ee Here ${\;}^{(\;\;)}$ means
symmetrization and $D^i_\alpha, \bar D^k_{\dot\alpha}$ are $N=2$ spinor
covariant derivatives satisfying the relations \be \{D^i_\alpha, D^k_\beta\} =
\{\bar D^k_{\dot\alpha}, \bar D^i_{\dot\beta}\} = 0\,, \quad \{D^i_{\alpha},
\bar D^k_{\dot\beta}\} = 2i\epsilon^{ik}
(\sigma^m)_{\alpha\dot\beta}\partial_m\,.\lb{N2com} \ee Using \p{N2com}, it is
a direct exercise to check that \p{HypC} gives rise to the equations of motion
for the physical component fields in $q^i = f^i + \theta^{i\alpha}\psi_\alpha +
\bar\theta^i_{\dot\alpha}\bar\chi^{\dot\alpha} + \ldots\,$, viz., \be \Box f^i
= 0\,, \quad \partial_m\psi \sigma^m = \sigma^m\partial_m\bar\chi =
0\,.\lb{MassHyp} \ee This phenomenon is a reflection of the ``no-go'' theorem
\cite{nogo} stating that no off-shell representation for hypermultiplet in its
``complex form'' (i.e. with bosonic fields arranged into $SU(2)$ doublet) can
be achieved with any {\it finite} number of auxiliary fields. It remained to
explore whether there exists a reasonable way to evade this theorem and to
write a kind of off-shell action for the hypermultiplet.

It was as well unclear how to construct a geometric unconstrained formulation
of the $N=2$ SYM theory, similar to the prepotential formulation of $N=1$ SYM.
The differential geometry constraints defining this theory were given in
\cite{N2SYM} \be \{{\cal D}^{(i}_\alpha, {\cal D}^{k)}_\beta\} = \{\bar{\cal
D}^{(k}_{\dot\alpha}, \bar{\cal D}^{i)}_{\dot\beta}\} = \{{\cal
D}^{(i}_{\alpha}, \bar{\cal D}^{k)}_{\dot\beta}\} = 0\,,  \lb{N2symC} \ee where
${\cal D}_\alpha^i = D^i_\alpha + i {\cal A}^i_\alpha$ is a gauge-covariantized
spinor derivative. Luca Mezincescu was the first to find the solution of these
constraints in the Abelian case through an unconstrained prepotential
\cite{LMe}. However the latter possesses a non-standard dimension -2, and the
corresponding gauge freedom does not admit a geometric interpretation. So it
remained to see whether something like a nice  geometric interpretation of the
$N=1$ SYM gauge group and prepotential $V$ can be revealed in the $N=2$ case
(and higher $N$ cases). The same problem existed for superfield $N=2$ SG.

In \cite{GAn} Galperin, Ogievetsky, and me observed that extended SUSY, besides
standard chiral superspaces generalizing the $N=1$ one, also admit some other
type of invariant subspaces which we called ``Grassmann-analytic''. Like in the
case of chiral superspaces, these subspaces are revealed by passing to some new
basis in the general superspace, such that spinor covariant derivatives with
respect to some fraction of Grassmann variables become ``short'' in it. Then
one can impose Grassmann Cauchy-Riemann conditions with respect to these
variables, with preserving full SUSY. In the $N=2$ case, allowing the $U(2)$
automorphism symmetry to be broken down to $O(2)$, and making the appropriate
shift of $x^m$, one can define the complex ``$O(2)$ analytic subspace'' \be
(\tilde{x}^m\,, \;\theta^1_\alpha + i\theta^2_\alpha\,,
\;\bar\theta^1_{\dot\alpha} + i\bar\theta^2_{\dot\alpha})\,, \lb{O2Anal} \ee
which is closed under $N=2$ SUSY, and the related Grassmann-analytic
superfields. It was natural to assume that this new type of analyticity plays a
fundamental role in extended SUSY, similarly to chirality in the $N=1$ case. In
\cite{Anat} we found that the hypermultiplet constraints \p{HypC} imply that
different components of $N=2$ superfield $q^i$ ``live'' on different
$O(2)$-analytic subspaces. Since \p{HypC} is $SU(2)$ covariant, it was tempting
to ``$SU(2)$- covariantize'' the $O(2)$ analyticity.

All these problems were solved with invention of the harmonic superspace \cite{HSS}-\cite{book}.

\noindent{\bf 4.2 N=2 harmonic superspace.}
$N=2$ harmonic superspace (HSS)
is defined as the product
\be
(x^m, \theta_{\alpha\;i},
\bar\theta_{\dot\beta}^k )\otimes S^2\,. \lb{HSS}
\ee
Here, $S^2 \sim SU(2)_A/U(1)$, with $SU(2)_A$ being
the automorphism group of the $N=2$ superalgebra.
The internal 2-sphere $S^2$ is represented in
a parametrization-independent way by the lowest (isospinor) $SU(2)_A$
harmonics
\be
S^2 \in (u^+_i, u^-_k), \quad u^{+i}u_i^- =1, \quad u^{\pm}_i \rightarrow
\mbox{e}^{\pm i\lambda}u^{\pm}_i~.
\ee
It is assumed that nothing depends on the $U(1)$ phase $\mbox{e}^{i\lambda}$,
so one effectively deals with the 2-sphere $S^2 \sim SU(2)_A/U(1)$.
The superfields given on \p{HSS} (harmonic $N=2$ superfields)
are assumed to be expandable into the harmonic series on $S^2$,
with the set of all symmetrized products of $u^+_i, u^-_i$
as the basis. These series are fully specified by the $U(1)$ charge
of the given superfield.

The main advantage of HSS is the existence of an invariant subspace in it, the
$N=2$ analytic HSS with half of the original odd coordinates \bea
&& \left(x^m_A, \theta^+_\alpha, \bar\theta^+_{\dot\alpha},
u^{\pm i}\right)  \equiv \left(\zeta^M, u^{\pm}_i  \right)~, \label{22} \\
&& x^m_A = x^m -2i\theta^{(i}\sigma^m\bar\theta^{k)}u^+_iu^-_k~,
\quad \theta^+_\alpha = \theta_{\alpha}^iu^{+}_i~, \;
\bar\theta^+_{\dot\alpha} = \bar\theta_{\dot\alpha}^iu^{+}_i~. \nonumber
\eea
It is $SU(2)$ covariantization of the $O(2)$ analytic superspace \p{O2Anal}.
It is closed under $N=2$ SUSY transformations and is real with respect to
the special involution which is the product of the ordinary complex
conjugation and the antipodal map (Weyl reflection) of $S^2$. All $N=2$
supersymmetric theories have off-shell formulations in terms of unconstrained
superfields given on \p{22}, the {\it Grassmann analytic}
$N=2$ superfields.

\noindent\underline{\it N=2 Matter} is represented by $n$ hypermultiplet
superfields $q^{+}_{a} (\zeta, u)$ ($\overline{(q^{+}_{a})} =
\Omega^{ab}q^{+}_{b}~,\;$ $\Omega^{ab} = -\Omega^{ba}$; $a,b = 1, \dots 2n$ )
with the following general off-shell action: \be  \label{33} S_q = \int du
d\zeta^{(-4)} \left\{q^{+}_{a}D^{++}q^{+a} + L^{+4}(q^+, u^+, u^-) \right\}~.
\ee Here, $du d\zeta^{(-4)}$ is the appropriate (charged!) measure of
integration over the analytic superspace \p{22}, $D^{++} =
u^{+\;i}\frac{\partial}{\partial u^{-i}} - 2i \theta^+\sigma^m\bar\theta^+
\frac{\partial}{\partial x^m}$ is the analytic basis form of one of three
harmonic derivatives one can define on $S^{2}$ (it is distinguished in that it
preserves the harmonic Grassmann analyticity) and the indices are raised and
lowered by the $Sp(n)$ totally skew-symmetric tensors $\Omega^{ab},
\Omega_{ab}$, $\Omega^{ab}\Omega_{bc} = \delta^a_c $. The interaction
Lagrangian $L^{+4}$ is an arbitrary function of its arguments, the only
restriction is its harmonic $U(1)$ charge $+4$ which is needed for the whole
action to be neutral. The crucial feature of the general $q^+$ action \p{33} is
an infinite number of auxiliary fields coming from the harmonic expansion on
$S^2$. This allowed one to circumvent the no-go theorem about the non-existence
of off-shell formulations of the $N=2$ hypermultiplet in its complex form. The
on-shell constraints \p{HypC} (and their nonlinear generalizations) amount to
{\it both} the harmonic analyticity of $q^+$ (which is a kinematic property
like $N=1$ chirality) and the dynamical equations of motion following from the
action \p{33}. After eliminating infinite sets of auxiliary fields by their
equations of motion, one gets the most general self-interaction of $n$
hypermultiplets. It yields in the bosonic sector the generic sigma model with
$4n$-dimensional hyper-K\"ahler (HK) target manifold in accord with the theorem
of Alvarez-Gaum\'e and Freedman about the one-to-one correspondence between
$N=2$ supersymmetric sigma models and HK manifolds \cite{AGF}. In general, the
action \p{33} and the corresponding HK sigma model possess no any isometries.
The object $L^{+4}$ is the HK potential, analog of the K\"ahler potential of
$N=1$ supersymmetric sigma models: taking one or another specific $L^{+4}$, one
gets the explicit form of the relevant HK metric after eliminating auxiliary
fields from \p{33}. So the general hypermultiplet action \p{33} provides an
efficient universal tool of the {\it explicit} construction of the HK metrics.

\noindent\underline{\it N=2 super Yang-Mulls theory} has as its fundamental
geometric object the analytic harmonic connection $V^{++}(\zeta,u)$ which
covariantizes the analyticity-preserving harmonic derivative: \bea D^{++}
\rightarrow {\cal D}^{++} = D^{++} +ig V^{++}~, \quad (V^{++})' = {1\over ig}
\mbox{e}^{i\omega}\left( D^{++} + ig V^{++} \right)\mbox{e}^{-i\omega}~, \eea
where $g$ is a coupling constant and $\omega(\zeta,u)$ is an arbitrary analytic
gauge parameter containing infinitely many component gauge parameters in its
combined $\theta, u$-expansion. The harmonic connection $V^{++}$ contains
infinitely many component fields, however almost all of them can be gauged away
by $\omega(\zeta,u)$. The rest of the $(8+8)$ components is just the off-shell
content of $N=2$ vector multiplet. More precisely, in the WZ gauge $V^{++}$ has
the following form:
\bea V^{++}_{WZ} &=& (\theta^+)^2 w(x_A) + (\bar\theta^+)^2
\bar{w}(x_A) + i\theta^+\sigma^m\bar\theta^+ V_m(x_A)
+ (\bar\theta^+)^2\theta^{+\alpha} \psi_{\alpha}^i(x_A)u^-_i \nn \\
&+& (\theta^+)^2\bar\theta^+_{\dot\alpha}\bar\psi^{\dot\alpha i}(x_A)u^-_i +
(\theta^+)^2(\bar\theta^+)^2 D^{(ij)}(x_A)u^-_iu^-_j ~. \eea
Here, $V_m, w,\bar
w, \psi^{\alpha}_i, \bar\psi^{\dot\alpha i}, D^{(ij)}$ are the gauge field,
complex physical scalar field, doublet of gaugini and the triplet of auxiliary
fields, respectively. All the geometric quantities of the $N=2$ SYM theory
(spinor and vector connections, covariant superfield strengths, etc), as well
as the invariant action, admit a concise representation in terms of
$V^{++}(\zeta,u)$. In particular, the closed $V^{++}$ form of the $N=2$ SYM
action was found in \cite{ZUP1}.

\noindent\underline{\it N=2 conformal supergravity} (Weyl) multiplet is represented in HSS
by the analytic vielbeins covariantizing $D^{++}$ with respect to
the analyticity-preserving diffeomorphisms of the superspace
$\left(\zeta^M, u^{\pm i}\right)$:
\bea
&&D^{++} \rightarrow {\cal D}^{++} = u^{+\,i}\frac{\partial}{\partial u^{-i}}
+ H^{++\, M}(\zeta, u)\frac{\partial}{\partial \zeta^M} +
H^{++++}(\zeta, u)u^{-i}\frac{\partial}{\partial u^{+ i}}~,  \nn \\
&& \delta \zeta^M = \lambda^M(\zeta, u)~, \quad \delta u^{+}_i =
\lambda^{++}(\zeta, u)u^{-}_i~, \nn \\
&& \delta H^{++\,M} = {\cal D}^{++}\lambda^M -
\delta^{M}_{\mu+}\theta^{\mu+}\lambda^{++}~, \quad \delta H^{++++}
= {\cal D}^{++}\lambda^{++}~, \mu \equiv (\alpha, \dot\alpha)~, \nn \\
&&\delta {\cal D}^{++} = -\lambda^{++} D^0~, \quad D^0 \equiv
u^{+i}\frac{\partial}{\partial u^{+i}} - u^{-i}\frac{\partial}{\partial u^{-i}}
+ \theta^{\mu+} \frac{\partial}{\partial \theta^{\mu+}}~. \eea The vielbein
coefficients $H^{++M}, H^{++++}$ are unconstrained analytic superfields
involving an infinite number of the component fields which come from the
harmonic expansions. Most of them, like in $V^{++}$, can be gauged away by the
analytic parameters $\lambda^M, \lambda^{++}$, leaving in the WZ gauge just the
$(24+24)$ $N=2$ Weyl multiplet. The invariant actions of various versions of
{\it $N=2$ Einstein} SG are given by a sum of the action of $N=2$ vector
compensating superfield $H^{++ 5}(\zeta, u), \delta H^{++ 5} = {\cal
D}^{++}\lambda^5(\zeta, u)\,$, and that of matter compensator superfields, both
in the background of $N=2$ conformal SG. The superfield $H^{++ 5}(\zeta, u)$
and extra gauge parameter $\lambda^5(\zeta,u)$ have, respectively, the
geometric meaning of the vielbein coefficient associated with an extra
coordinate $x^5$ (central charge coordinate) and the shift along this
coordinate. Nothing is assumed to depend on this coordinate. The most general
off-shell version of $N=2$ Einstein SG is obtained by choosing the superfield
$q^{+a}(\zeta, u)$ as the matter compensator. It involves an infinite number of
auxiliary fields and yields all the previously known off-shell versions with
finite sets of auxiliary fields via appropriate superfield duality
transformations. Only this version allows for the most general SG-matter
coupling. The latter gives rise to a generic quaternion-K\"ahler sigma model in
the bosonic sector, in accordance with the theorem of Bagger and Witten
\cite{BW}.

More references to the HSS-oriented works of the Dubna group can be found
in the book \cite{book}.

\vspace{0.3cm}

\noindent{\bf 4.3 Some further developments.} Here we sketch a few  basic
directions in which the HSS method was developed after its invention in
\cite{HSS}. It can be generalized to $N\geq 2$. It was used to construct, for
the first time, an unconstrained off-shell formulation of the $N=3$ super YM
theory (equivalent to $N=4$ YM on shell) in the harmonic $N=3$ superspace with
the purely harmonic part $SU(3)/[U(1)\times U(1)]$, $SU(3)$ being the
automorphism group of $N=3$ SUSY \cite{4}. The corresponding action is written
in the analytic $N=3$ superspace and has a nice form of the superfield
Chern-Simons term. The $N=4$ HSS with the harmonic part $SU(4)/[U(1)\times
SU(2)\times SU(2)]$ was employed to give a new geometric interpretation of the
on-shell constraints of $N=4$ super YM theory \cite{5}. In \cite{IS,I1} the
bi-harmonic superspace with two independent sets of $SU(2)$ harmonics was
introduced and shown to provide an adequate off-shell description of $N=(4,4),
2D$ sigma models with torsion. $N=4$, $1D$ HSS was used in \cite{DI} to
construct a new super KdV hierarchy, $N=4$ supersymmetric one. Various versions
of HSS in diverse dimensions were also explored in \cite{ZU2}. The current important
applications of the HSS approach involve the quantum off-shell calculations in
$N=2$ and $N=4$ gauge theories (see, e.g., \cite{31,32}), classifying ``short''
and ``long'' representations of various superconformal groups in diverse
dimensions in the context of the AdS/CFT correspondence \cite{33}, uses in
extended supersymmetric quantum mechanics models \cite{IL,BIS}, study of the
domain-wall solutions in the hypermultiplet models \cite{AIN}, description of
self-dual supergravities \cite{DO}, etc. The Euclidean version of $N=2$ HSS was
applied in \cite{ILZ,FILSZ,ILZ2} to construct string theory-motivated
non-anticommutative (nilpotent) deformations of $N=(1,1)$ hypermultiplet and
gauge theories. Recently, using the HSS approach, the first example of
renormalizable $N=(1,0)$ supersymmetric $6D$ gauge theory was constructed \cite{ISZ}.

No doubt the HSS method as the most appropriate approach to off-shell theories
with extended supersymmetries will be widely used and advanced in future
studies including those to be carried out in Dubna.

\section{Other SUSY-related activities}
Besides the mainstream SUSY researches outlined in the previous Sections, there were several
important pioneering achievements of Dubna group in the fields related to some other
applications of supersymmetry.

First of all, these are the issues related to two-dimensional supersymmetric
integrable systems. In \cite{N2L}, together with S.O. Krivonos, we constructed
an integrable $N=(2,2)$ extension of the Liouville theory which was unknown
before. The first example of $N=(4,4)$ integrable system, the $N=(4,4)$
WZW-Liouville theory \footnote{WZW (Wess-Zumino-Witten) stands for sigma models
on group manifolds with torsion.}, was presented in \cite{IK1} and further
studied (at the classical and quantum levels) in \cite{IKL1,IKL2}. In
\cite{IK1} we, independently of the authors of \cite{GHL}, discovered $N=(4,4)$
twisted multiplet and in fact gave the first example of supersymmetric WZW
model (at once with $N=(4,4)$ supersymmetry). Superfield actions of $N=4$ and
higher $N$ supersymmetric and superconformal quantum mechanics were pioneered
in our papers \cite{IKL3,IKP,ISmi}. New integrable super KdV and NLS type
hierarchies were discovered in \cite{DI,IK2,Bus,KTS,BKS,IKT}. The manifestly
$N=2$ supersymmetric superfield Hamiltonian reduction as a powerful method of
constructing $N=2$ super $W$ algebras and integrable systems was developed in
\cite{Hal}.

Interesting exercises in the superstring theory were undertaken in the unpublished papers \cite{II1}, following
ref. \cite{II2}. There, we considered a generalization of the standard Green-Schwarz superstring
to certain supergroup manifolds using the powerful techniques of Cartan 1-forms, found the conditions
for kappa invariance of the relevant curved actions (with specific non-trivial examples), constructed
Hamiltonian formalism for these models and showed their classical integrability. It is curious that
this study was fulfilled about 10 years prior to emerging the current vast interest in such constructions
within the AdS/CFT paradigm.

Another, more recent activity associated with superbranes was related to their
superfield description as systems realizing the concept of Partial Breaking of
Global Supersymmetry (PBGS) pioneered by Bagger and Wess \cite{BW1} and Hughes
and Polchinsky \cite{HP}. In this approach, the physical worldvolume superbrane
degrees of freedom are accommodated by Goldstone superfields, on which the
worldvolume SUSY is realized by linear transformations and so is manifest. The
rest of the full target SUSY is realized nonlinearly, a la Volkov-Akulov. In
components, the corresponding Goldstone superfield actions yield a static-gauge
form of the relevant Green-Schwarz-type worldvolume actions. In the cases when
Goldstone supermultiplets are vector ones, the Goldstone superfield actions
simultaneously provide appropriate supersymmetrizations of the Born-Infeld
action. The references to works of the Dubna group on various aspects of the
PBGS approach and superextensions of the Born-Infeld theory can be found, e.g.,
in the review papers \cite{I,II}. Among the most sound results obtained on this
way I would like to distinguish the interpretation of the hypermultiplet as a
Goldstone multiplet supporting partial breaking of $N=1,\, 10D$ SUSY
\cite{10susy}, the construction of $N=2$ extended Born-Infeld theory with
partially broken $N=4$ SUSY \cite{BIK,N4N2}, as well as of $N=3$ superextension
of Born-Infeld theory with the use of the $N=3$ HSS approach \cite{N3}. Closely
related issues of the twistor-harmonic description of superbranes in diverse
dimensions were addressed in \cite{IKaSB,DIS}. Recently, the superfield PBGS
approach was generalized to partially broken AdS supersymmetries (see
\cite{Ilast} and refs. therein).

One of the current research activities is the study of models of supersymmetric
quantum mechanics with extended $N=4$ and $N=8$ SUSY. It continues and advances
the directions initiated by the papers \cite{IKL3,IKP,ISmi} and aims at further
understanding of the structure of the parent higher-dimensional supersymmetric
field theories, as well as of the AdS$_2$/CFT$_1$ version of general
string/gauge correspondence. For some of the latest developments in this area
see, e.g., \cite{BILK1,IL,BIS,Nonl1,Nonl2}.

At last, let me mention recent papers \cite{IMT1} which treat supersymmetric versions of the
quantum-mechanical Landau problem on a plane and two-sphere, as well as the closely related issue of ``fuzzy''
supermanifolds (which are non-anticommutative versions of the classical supermanifolds, such that
their superspace coordinates form a superalgebra isomorphic to that of superisometries of the
classical supermanifold). This direction of research looks very prospective from the physical
point of view, since it is expected to give rise to a deeper understanding of quantum Hall effect and
superextensions thereof, equally as of the relationships of these models with superparticles and
superbranes.

\end{document}